\documentclass[10pt]{article}
\usepackage[margin=1in]{geometry}

\usepackage{times}
\usepackage{amsmath}
\usepackage{latexsym}
\usepackage{amsfonts}
\usepackage{amssymb}
\usepackage{amsthm}

\theoremstyle{definition}
\newtheorem{theorem}{Example}{\itshape}{\normalfont}

\usepackage{nicefrac} 
\usepackage{mathtools} 

\usepackage[misc]{ifsym} 

\usepackage{xspace,xcolor}
\usepackage{paralist}
\usepackage{enumitem}
\usepackage{siunitx}
\DeclareSIUnit\knot{kn}

\usepackage{acronym}

\usepackage[pdftex]{graphicx}

\newcommand{\eg}{e.g.,\xspace}
\newcommand{\ie}{i.e.,\xspace}

\newcommand{\acate}{AdvoCATE\xspace}

\newcommand{\spi}[1]{\mathcal{I}_{\mathtt{#1}}}
\newcommand{\metric}[1]{\mathtt{#1}}
\newcommand{\eventid}[1]{\mathsf{E}_{#1}}
\newcommand{\barrierid}[1]{\mathsf{B}_{#1}}
\newcommand{\rr}{\mathtt{RR}}

\acrodef{smb}[SMB]{safety measurement basis}
    
\usepackage{hyperref}
\hypersetup{
    colorlinks=true,	
    linkcolor=blue,	
    citecolor=blue,	
    filecolor=blue,    
    urlcolor=blue		
}

\usepackage{fancyhdr}
\fancypagestyle{firstpage}
{
    \fancyhead[C]{\textcolor{red}{The content of this author pre-print version has been accepted for publication after peer review but is not the Version of Record. The Version of Record appears in the Proceedings of the 43nd International Conference on Computer Safety, Reliability, and Security (SAFECOMP 2024), Florence, Italy.}}   
    }

\begin{document}


\title{Reconciling Safety Measurement and Dynamic Assurance\thanks{%
This work was performed under Contract No.~80ARC020D0010 with the National Aeronautics and Space Administration (NASA), with support from the System-wide Safety project, under the Airspace Operations and Safety Program of the NASA Aeronautics Research Mission Directorate.
The United States Government retains and the publisher, by accepting the article for publication, acknowledges that the United States Government retains a non-exclusive, paid-up, irrevocable, worldwide license to reproduce, prepare derivative works, distribute copies to the public, and perform publicly and display publicly, or allow others to do so, for United States Government purposes. All other rights are reserved by the copyright owner.
}}

\author{Ewen Denney and Ganesh Pai\\
			\normalsize{KBR / NASA Ames Research Center, 
						Moffett Field, CA 94035, USA}\\
			\normalsize{\{ewen.denney, ganesh.pai\}@nasa.gov}
		}

\date{}

\maketitle

\thispagestyle{firstpage}

\begin{abstract}
We propose a new framework to facilitate dynamic assurance within a safety case approach by associating safety performance measurement with the core assurance artifacts of a safety case. The focus is mainly on the \emph{safety architecture}, whose underlying risk assessment model gives the concrete link from safety measurement to operational risk. Using an aviation domain example of autonomous taxiing, we describe our approach to derive safety indicators and   revise the risk assessment based on safety measurement. We then outline a notion of \emph{consistency} between a collection of safety indicators and the safety case, as a formal basis for implementing the proposed framework in our tool, \acate.  
\end{abstract}

\section{Introduction}\label{s:introduction}

Software-based self-adaptation and machine learning (ML) technologies for enabling autonomy in complex systems---such as those in civil aviation---may induce new and unforeseen ways for operational safety performance to deviate from an approved baseline of acceptable risk. This phenomenon, known as \emph{practical drift}~\cite{icao-smm}, emerges from the inevitable variabilities in real-life operations to meet service expectations in an operating environment that is inherently dynamic. 
Conceptually, it can be understood as progressively imperceptible reductions in the safety margins built into a system in part due to initially benign operational tradeoffs between safety and performance. 
A system therefore appears to be operating safely but, in fact, is operating at a higher level of safety risk than what was originally considered acceptable, or approved for service. Left unchecked, practical drift may suddenly manifest as a serious \emph{incident} or \emph{accident}. Assessing the change in operational safety risk is thus key to identifying practical drift, its impact, and the mitigations needed.

\subsection{Related Work}

The conventional approach to operational safety assurance in aviation largely relies upon hazard tracking and safety performance monitoring and measurement, as part of a larger \emph{safety management system} (SMS)~\cite{faa-ato-sms}. 
The contemporary safety case approach to assurance has similarly employed safety monitoring and measurement: for example, our earlier work on \emph{dynamic safety cases}~\cite{dhp-icse-nier-2015} first suggested connecting safety monitoring to assurance argument modification actions. 
Subsequently, an approach to defining performance metrics and monitors by identifying the defeaters and counterarguments to a safety case has been developed in~\cite{hawkins2023}. The concept has since also been applied to safety assurance of self-adaptive software~\cite{calinescu-tse-2018}, and to detect operational exposure to previously unknown hazardous conditions~\cite{schleiss-icsrs22}.
More recently, the use of \emph{safety performance indicators} (SPIs)---a concept with a well-established history of use in aviation safety~\cite{icao-smm}---has been proposed for evaluating safety cases for autonomous vehicles~\cite{koopman-book}.
These approaches all share a common motivation: using measurement based assessment to confirm at deployment, and maintain in operation, the validity of the assurance arguments of a safety case. 

Although such an approach suggests which parts of an argument may have been invalidated, and thus require changing, the nature and extent of the change to operational safety risk levels is left implicit. Such analyses can also meaningfully inform what modifications may be needed to the system and its safety case, especially when---due to practical drift---improved system performance is observed without detrimental safety effects, even though parts of the safety argument have become invalid.
Current safety case approaches that use safety performance measurement to validate assurance arguments give limited guidance on how to facilitate what this paper considers as \emph{dynamic} assurance (see Fig.~\ref{f:framework}): \emph{continued, justified confidence that a system is operating at a safety risk level consistent with an approved risk baseline}. 

There are other variations of the dynamic assurance concept~\cite{sinadra}, \cite{trapp2019} that aim to optimize operational system performance, and thus opt for situation-specific runtime tradeoffs between safety and functional performance, instead of designing for the worst case. 
However, such tradeoffs may result in the initiating conditions for practical drift. Our proposed framework rather aims to identify and contain practical drift, whilst considering that a safety case for a system is always for a design that accounts for the worst credible safety effects. 
In~\cite{schleiss-icsrs22}, dynamic assurance refers to the automated aspects of so-called \emph{continuous assurance}: a concept that, in effect, extends our prior work~\cite{dhp-icse-nier-2015}, by using monitors for different kinds of uncertainty that then trigger modifications to the system and its assurance case.
The relationship of testing and operational metrics to safety assurance has been explored in~\cite{strigini-ivds21}, similar to our work in this paper (see Section~\ref{s:framework}), though there the focus is on providing confidence that a system meets its safety target, given evidence of mishap-free operation. In contrast, our focus here is on determining how safety risk has changed given similar measurement evidence.

\subsection{Contributions and Paper Organization}

To facilitate a framework for dynamic assurance within a safety case approach, the focus of this paper is on associating safety performance measurement with the \emph{safety architecture} of a system, in addition to assurance arguments (Section~\ref{s:background}). 
Using an aviation domain system (Section~\ref{s:example}) as motivation, we present our approach to define safety metrics and indicators, through a concept of \emph{safety measurement basis} (SMB), then revise the operational safety risk assessment based on safety measurement, and characterize the change to safety risk levels (Section~\ref{s:framework}). Additionally, we give illustrative numerical examples. Then (Section~\ref{s:formal-foundations}) we formalize a notion of \emph{consistency} between the SMB for a system and the arguments of its safety case. We conclude (Section~\ref{s:conclusions}) by describing a preliminary implementation in \acate, and with a discussion of our future plans to further advance this work. The contributions above differentiate our work from prior related research.

\section{Conceptual Background}\label{s:background}

\subsection{Safety Case Metamodel}\label{ss:advocate-bg}

Our safety case concept~\cite{dac-computer} communicates confidence in safety through multiple viewpoints via a collection of core, interlinked \emph{assurance artifacts}, namely: hazard, requirement, and evidence logs, a safety architecture, and an assurance rationale. Of those, the last two are particularly relevant for this paper. Assurance rationale captured as structured arguments expresses the reasoning why safety claims ought to be accepted on the basis of the evidence supplied. A safety architecture~\cite{djp-dasc2018}, \cite{dpw-ress2019} models the mitigations (and their interrelations) to the events characterizing the operational \emph{risk scenarios} for a system, thereby offering a system-level viewpoint on how safety risk is reduced.

\begin{figure}[t]
	\centering
	\includegraphics[width=\columnwidth]{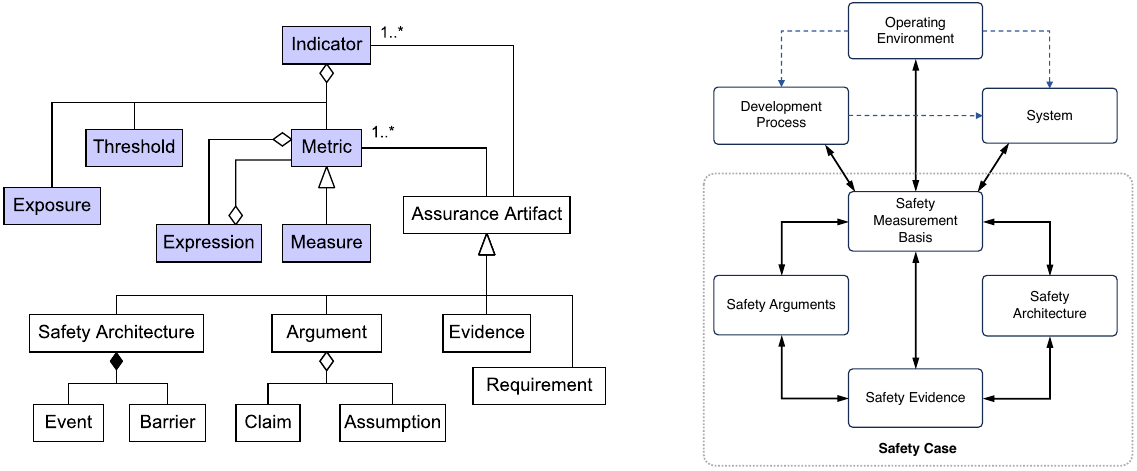}
	\caption{Fragment of \acate safety case metamodel (on the left) extended with measurement concepts that are part of a safety measurement basis (shown on the right), which is the interface between quantities in the system, its environment, its development process, and the assurance artifacts comprising a safety case (solid arrows denote consistency relations).} 
	\label{f:dynamic-metamodel}
\end{figure}

Fig.~\ref{f:dynamic-metamodel} shows a fragment of the metamodel associated with our safety case concept (as unshaded class nodes), for which we have a model-based implementation in our tool, \acate~\cite{dp-jase2017}. We use the \emph{goal structuring notation} (GSN)~\cite{gsn-std-v3} to represent structured arguments, and \emph{bow tie diagrams} (BTDs) to represent views of a safety architecture. Those views capture a causal chain (\eg see Fig.~\ref{f:btd-example}) of \emph{threats} (initiating events) causing a \emph{top event} (a hazard) that can lead to \emph{consequence events} (undesired safety effects), along with the \emph{barriers} (mitigations) necessary to reduce the safety risk posed. Each such event chain requires a combination of \emph{hazardous activity}, \emph{environmental condition}, and \emph{system state} (together representing the operating context\footnote{Also known as an \emph{operational design domain} (ODD) for systems integrating ML~\cite{kape-safecomp-2023}.}), and can admit an arbitrary number of intermediate events between the initiating threat and terminating consequence events. 
Each barrier is itself a system comprising underlying \emph{controls}; thus, it can have its own associated safety architecture, giving the overall model a layered structure that can mirror the system hierarchy.

A risk assessment model underlying a safety architecture gives the formal basis to:
\begin{inparaenum}[(i)]
	\item characterize the extent of risk reduction, and 
	\item link safety metrics and indicators to operational safety risk 
	(see Section~\ref{s:framework}). 
\end{inparaenum}
In brief, this model relates the risk of consequence events, \ie their probability and severity, to that of the precursor events, and to the \emph{integrity}\footnote{Integrity is the probability that a barrier or control is not breached, \ie it delivers its intended function for reducing risk in the specified operating context and scenario~\cite{dpw-ress2019}.} of the applicable barriers and their constituent controls. Depending on the stage of system development, we can interpret each of an event probability and barrier/control integrity both as a design target and verification goal. For the rest of this paper, we mainly consider the risk reduction contribution of barriers.

\subsection{Safety Measurement}\label{ss:safety-measurement}

We extend the safety case metamodel in \acate with concepts for safety performance measurement (shown by the shaded class nodes in Fig.~\ref{f:dynamic-metamodel}) as follows: we link the \emph{indicators} to the core assurance artifacts---in particular, the event and barrier elements of a safety architecture, the claims and assumptions in arguments, to requirements, and to evidence artifacts.
An \emph{indicator} consists of a \emph{metric} along with a \emph{threshold}, representing the target that a metric should (or should not) reach, over a specified \emph{exposure}, expressed either as a duration of continuous time or a specified number of occurrences of a discrete event. Indicators that have a bearing on safety can be called \emph{safety indicators} (SIs) or \emph{safety performance indicators} (SPIs). \emph{Metrics} are computed values based on \emph{measures}---directly observable parameters of the system, its environment, and its development process---and other metrics, which we represent using an expression language. Thus, they are arithmetic expressions over measured variables drawn from the most recent \emph{mission}---which we term as a \emph{data run}---or the missions conducted over the lifetime of the system. They can also refer to values referenced in assurance artifacts. 

\acused{smb}

As shown in Fig.~\ref{f:dynamic-metamodel}, a safety case can be seen as comprising a \emph{dynamic} portion (indicators, metrics, and measures) and a static portion (safety arguments and safety architecture), with links associating the two. We refer to the set of interconnected indicators, metrics, and measures, along with their traceability links to the assurance artifacts of a safety case as a \emph{safety measurement basis} (\ac{smb}). 
Roughly speaking, the connection between the dynamic and static portions is that the indicators represent the objectively quantifiable content of the arguments and the safety architecture which, in turn, give the justification for how those indicators collectively provide safety substantiation. 
Put another way, we want the \ac{smb} to be \emph{consistent} with the static portions of the safety case, especially the arguments and the safety architecture (see Section~\ref{s:formal-foundations}). 

\section{Motivating Example}\label{s:example} 

We motivate this work using an aviation domain use case of autonomous aircraft taxiing~\cite{dac-computer}. This system uses a \emph{runway centerline tracking} function comprising a classical controller coupled to a deep convolutional neural network that estimates aircraft position from optical sensor data. 
The functional objective is to maintain both the \emph{cross-track error} (CTE) and the \emph{heading error} (HE) within pre-defined bounds. CTE is the horizontal distance between the runway centerline and the aircraft body (or \emph{roll}) axis; HE is the angle between the respective headings of the runway centerline and the roll axis. The safety objective is to avoid a lateral \emph{runway overrun} (also known as a \emph{runway excursion}), \ie departing the sides of the runway.

\begin{figure}[t]
	\centering
	\includegraphics[width=\columnwidth]{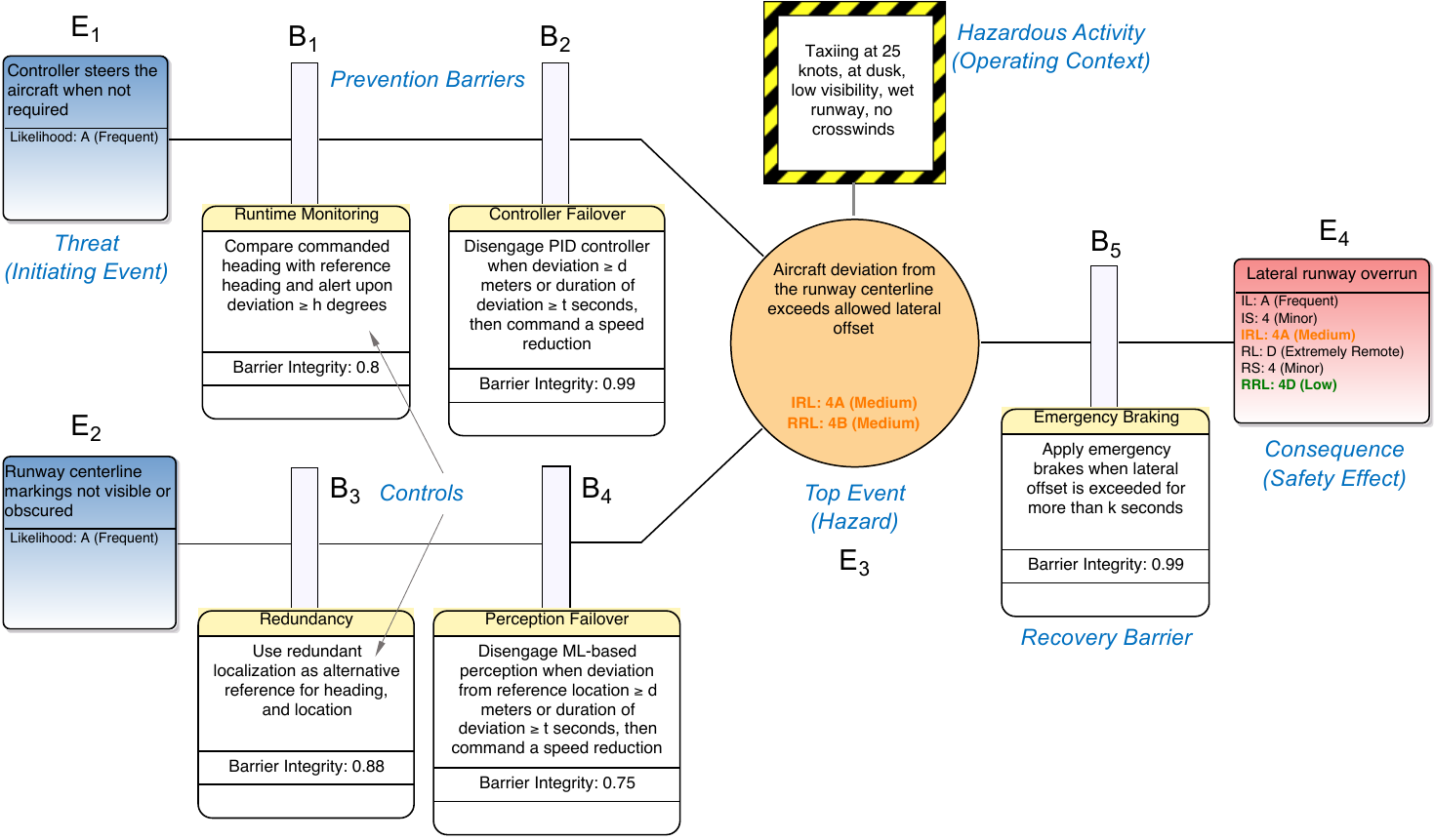}
	\caption{Annotated BTD fragment for an autonomous taxiing capability, showing how a lateral runway overrun is mitigated under specific initiating events leading to centerline tracking violation.} 
	\label{f:btd-example}
\end{figure}

Fig.~\ref{f:btd-example} shows a BTD fragment for this example (annotated to show its graphical elements and their identifiers) as a view of its wider safety architecture (not shown), which composes~\cite{dpw-ress2019} similar such BTDs, albeit for different operating contexts, threats, top events, and consequences. Here, the operating context involves a relatively low speed (\SI{25}{\knot}), low visibility taxi operation on a wet runway, at dusk, under no crosswind conditions. The hazard to be controlled ($\eventid{3}$) is a violation of the allowed lateral offset from the runway centerline, failing which a lateral runway overrun ($\eventid{4}$) could occur. Two (out of many) initiating causes for this hazard have been shown: a controller malfunction that steers the aircraft away from the centerline when not required ($\eventid{1}$); and runway centerline markings that are not visible, or are obscured ($\eventid{2}$).

\subsection{Baseline Safety}\label{ss:baseline-safety}

To characterize the safety risk level of an operating scenario, we use the risk assessment model associated with the safety architecture to establish a \emph{baseline level} of operational safety risk for the identified safety effects. 

For the scenario in Fig.~\ref{f:btd-example}, the \emph{initial risk level} (IRL) of the consequence event $\eventid{4}$ is labeled $\mathsf{4A (Medium)}$. That is, $\eventid{4}$ has a \emph{medium} level of unmitigated risk, and is assigned the \emph{risk classification category} $\mathsf{4A}$. That refers to a region of the overall risk space that has been discretized using a classical $5 \times 5$ risk matrix of categories of consequence event probability, ranging from \emph{Frequent} (A) to \emph{Extremely Improbable} (E), and consequence event severity, ranging from \emph{Minimal} (5) to \emph{Catastrophic} (1). For a definition of those categories, see~\cite{faa-ato-sms}.
A similar interpretation applies to \emph{residual risk level} (RRL) which, for $\eventid{4}$, is shown as $\mathsf{4D (Low)}$, representing the risk remaining after mitigation using the indicated barriers and the associated controls. Specifically, $\barrierid{1}$: \emph{Runtime Monitoring}, $\barrierid{2}$: \emph{Controller Failover}, $\barrierid{3}$: \emph{Redundancy}, and $\barrierid{4}$: \emph{Perception Failover}, serve as prevention barriers for exceeding the allowed CTE, while $\barrierid{5}$: \emph{Emergency Braking} is a recovery barrier invoked after the top event occurs. 

For aeronautical applications, civil aviation regulations and the associated certification or approval processes generally establish what constitutes \emph{acceptable} and \emph{approved} baseline risk levels respectively. The two can be the same (though they need not be) and, typically, are given in terms of a so-called \emph{target level of safety} (TLOS), which specifies the (maximum acceptable) probability of the undesired safety effect per unit of operational exposure, \eg $10^{-6}$ lateral runway excursions per taxi operation.
How TLOS is established and approved is out of scope for this paper\footnote{Interested readers may refer to~\cite{tlos-tr}.}; as such, in Fig.~\ref{f:btd-example}, either of the values of the IRL, $\mathsf{4A (Medium)}$, or the RRL, $\mathsf{4D (Low)}$, may plausibly meet the TLOS, and therefore could be an approved baseline level of safety risk. For the purposes of this example, we assume that the RRL shown is the approved baseline that meets the TLOS. Once a system is deployed, note that the RRL for an event is, in fact, \emph{dynamic}, \ie as a sequence of values starting from the approved baseline, it represents how the risk of that event evolves over the system lifetime (also see Fig.~\ref{f:framework}).

\subsection{Practical Drift}\label{ss:practical-drift}

Some barriers or controls in the safety architecture of a system may be relaxed in operation to improve the performance of system services and/or to make local optimizations that address the operating context. 
In our running example, for instance, to increase runway throughput whilst operating on large runways in better environmental conditions (\eg clear weather, and dry runway surface), the time an aircraft spends on a runway could be reduced. For that purpose, suppose that disengagement of the perception function or the controller is delayed (see Fig.~\ref{f:btd-example}), or that more permissive CTE bounds are admitted. 
In those cases, the system may enter certain states that would have been prohibited otherwise. In particular, such states represent violations of the barrier/control requirements that were stated as claims in the pre-deployment safety case. 

However, when there is improved system performance without observed safety consequences or mishaps, those states are not perceived as violations that increase residual risk. This can lead to misplaced assurance in operational safety when the system as operated deviates from its safety case. Practical drift can then emerge when multiple safety mitigations may be progressively loosened, and continued, incident-free system operations under such changes obscure the increase in operational safety risk. It is important to emphasize that relaxing mitigations to improve performance represents an operational tradeoff rather than a deliberate attempt to subvert safety. An analogy, for example, is highway driving at the speed of traffic that exceeds the posted speed limits---a practice that is not always unsafe, but poses higher risk in general.

\section{Framework}\label{s:framework}

\begin{figure}[htb]
	\centering
	\includegraphics[width=0.72\textwidth]{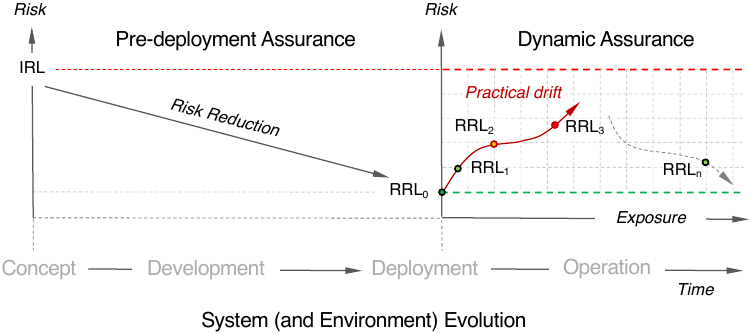}
	\caption{Pre-deployment assurance gives justified confidence in the reduction of an initial risk level (IRL) of the safety effects in a system concept, through system development, to a baseline residual risk level (RRL$_{0}$) that meets the TLOS at deployment. Dynamic assurance provides confidence in operation that the approved baseline is maintained, by identifying and managing practical drift.}
	\label{f:framework}
\end{figure}

Dynamic assurance within a safety case approach gives a proactive means to assess and contain practical drift through continued assurance that the operational safety risk level for the system is aligned with its approved baseline (see Fig.~\ref{f:framework}). A framework that enables this must at least:
\begin{inparaenum}[(i)]
	\item \label{ORL-change} characterize how operational safety risk levels have changed;
	\item determine which mitigations, if any, may be legitimately relaxed without safety deteriorating; and 
	\item identify the necessary modifications to both the system and its safety case, so that the two are mutually consistent during system operation.  
\end{inparaenum}
Next, we discuss how relating safety performance measurement to the safety architecture in a safety case, in addition to its arguments, gives the necessary elements and technical foundations for the first of the preceding three requirements---the main focus in this paper. The examples presented next are meant to be illustrative and not comprehensive.

\subsection{Defining Safety Metrics and Indicators}
\label{ss:safety-indicators}

A safety architecture and its associated risk assessment model~\cite{dpw-ress2019} give a basis to allocate safety targets to the safety functions, and subsequently confirm them (analytically and empirically).
TLOS is a system-level safety target always assigned to consequence event probability. Decomposing and allocating the TLOS across the elements of the safety architecture gives the safety integrity targets for barriers and controls, along with precursor event probabilities that we interpret as \emph{scenario-specific safety targets}. 
Relating safety targets to safety performance measurement in general, and safety indicators (SIs) in particular, facilitates tracking and confirming that the mitigations are performing in operation as intended.
One way to embed TLOS into an SI is by simply converting the corresponding probability value into an event frequency threshold applied to an appropriate 
safety metric used during development or in operation. In this section we focus on the operational safety metrics, addressing the metrics used during development in Section~\ref{ss:update}.

TLOS and the corresponding SIs can be 
\emph{generic}, \ie apply to all relevant operating contexts of a safety architecture, or \emph{scenario-specific}, \ie applicable to a particular operating context. 
For instance, let the TLOS for lateral runway overrun under all relevant operating conditions of the example system be $10^{-6}$ per taxi operation. We can then define a corresponding generic SI, $\spi{LRE}$: $\metric{opLatRwyEx} \leq 1$ in $10^{6}$ taxi operations, where $\metric{opLatRwyEx}$ is an operational safety metric\footnote{Henceforth, identifiers with the prefix `$\mathtt{dev}$' refer to metrics used during system development, and the prefix `$\mathtt{op}$' indicates an operational safety metric.} for the \emph{number of lateral runway overrun events in operation}, whose threshold value is $1$, measured over an exposure of $10^{6}$ taxi operations. 
Another commonly used unit of exposure is \emph{flight hours}~\cite{faa-srm-order}, and the SI can be given equivalently as $\spi{LRE}: \metric{opLatRwyEx} \leq 1$ in $10^{6}\times t$ flight hours, where $t$ is the average time in flight hours of a taxi operation.

Scenario-specific SI definition proceeds in the same way, but is applied to specific operating contexts after first decomposing and allocating the TLOS of a consequence event to its scenario-specific instance. For example, if $10\%$ of all taxi operations occur under the operating context of Fig.~\ref{f:btd-example}, then we can modify the exposure of $\spi{LRE}$ to $10^{5}$ taxi operations to get the scenario-specific SI for the consequence event $\eventid{4}$.

Similarly, we can define generic and scenario-specific SIs for the remaining safety architecture elements by converting the associated event probability and barrier integrity values as applicable. Moreover, recalling that a barrier can have its own safety architecture (Section~\ref{ss:advocate-bg}), we can iteratively define SIs for the lower layers of a system hierarchy. 
Thus, in Fig.~\ref{f:btd-example}, we can define the scenario-specific SI for the barrier $\barrierid{4}$: \emph{Perception Failover} as $\spi{PFO}$: $\metric{opPcpDisEngF} \leq y$ in $n$ taxi operations. In Section~\ref{ss:numerical-examples}, we illustrate one approach to instantiate $y$ and $n$. 

Here, $\metric{opPcpDisEngF}$ is a metric related to the integrity of $\barrierid{4}$ (itself a metric) that counts the \emph{number of failed disengagements of ML-based perception in operation}; its threshold value is $y$, to be measured over an exposure of $n$ taxi operations conducted in the stated operating context for the specified scenario. This metric relies upon a precise definition of a \emph{failed disengagement} (not given here), which may itself be given in terms of other metrics, \eg those associated with its \emph{functional deviations} (\ie violation of the requirements for the barrier, its constituent controls, or their verification), and its \emph{failure modes} (of the physical systems to which the barrier function is allocated).
Additional operational safety metrics related to barrier integrity include $\metric{opTxLowVisW}$, counting the \emph{number of taxi operations conducted at dusk under low visibility, no crosswind, and wet runway conditions} (\ie the operating context of Fig.~\ref{f:btd-example}), from which we may infer the \emph{number of successful disengagements of ML-based perception} as the metric $\metric{opPcpDisEngS} = \metric{opTxLowVisW} - \metric{opPcpDisEngF}$.

\subsection{Updating and Revising the Operational Risk Assessment}\label{ss:update}

A pre-deployment safety case represents what (we believe) a system design achieves at deployment, and will continue to achieve in operation. 
Some of the metrics and SIs applicable during system development constitute measurement evidence verifying safety performance, \eg during pre-deployment system testing or flight testing. Thus, by associating those metrics and SIs with the safety architecture, we get the \emph{prior} values of event probability and barrier integrity.
For the scenario and operating context of Fig.~\ref{f:btd-example}, some of the metrics used during system development for the barrier $\barrierid{4}$ are: $\metric{devTxLowVisW}$: the \emph{number of tests for $\barrierid{4}$} $= t$ (say); $\metric{devPcpDisEngS}$: the \emph{number of successful disengagements of ML-based perception} $= s$; and $\metric{devPcpDisEngF}$: the \emph{number of failed disengagements of ML-based perception} $= (t-s) = f$. 

If the test campaign during system development is designed as a Bernoulli process~\cite{ladkin2022} then we can model the sequence of test results as a binomial distribution, $\mathtt{Binom}\left(\chi: \eta, \theta \right)$, whose parameters are $\chi$: the number of successes, $\eta$: the number of independent trials, and $\theta$: the probability of success in each trial. 
Hence, we can assign the values of the metrics $\metric{devPcpDisEngS}$ and $\metric{devTxLowVisW}$, respectively, to the first two parameters as $\chi \coloneqq s$, and $\eta \coloneqq t$. Let $\theta \coloneqq p$, the unknown (fixed) probability that each test produces a successful disengagement. We can model $p$ as the conjugate prior beta distribution, $\pi(p) \sim \mathtt{Beta}\left(\alpha, \beta\right)$. The \emph{hyperparameters} (\ie the parameters of the prior distribution) represent our prior knowledge of the number of successful and failed tests during development. Hence we assign to them the values of the metrics $\metric{devPcpDisEngS}$ and $\metric{devPcpDisEngF}$, respectively, as $\alpha \coloneqq s$, and $\beta \coloneqq f$. The beta distribution mean, $\mu_{p} = \nicefrac{s}{t}$, gives a point estimate of the prior barrier integrity, and its variance, $\sigma_{p}^{2}=\nicefrac{sf}{t^{2}(t+1)}$, gives the uncertainty in that estimate.

In operation, safety performance measurement yields a sequence of observations of the state of the safety system. We can transform this data into a \emph{likelihood function}, \ie a joint probability of the observations given as a function of the parameters of a model of the underlying data generation process. 
In our example, a binomial probability density function (PDF) is a reasonable \emph{initial} model 
\begin{inparaenum}[(i)]
	\item assuming that the pre-deployment safety case provides the argument and evidence that testing is representative of actual operations (as would likely be necessary), and 
	\item since a binomial distribution models the sequence of test results.  
\end{inparaenum}	
Thus, supposing that over $n$ taxi operations conducted in the operating context of Fig.~\ref{f:btd-example}, there were $x$ failures to disengage ML-based perception on demand. We now have the operational safety metrics $\metric{opTxLowVisW} = n$, and $\metric{opPcpDisEngF} = x$, so that $\metric{opPcpDisEngS} = (n - x) = y$, and the likelihood function is $\mathcal{L}\left(p|n,y\right) = \binom{n}{y}\,p^{y}(1-p)^{x}$. As before, $p$ is the unknown probability of a successful disengagement of ML-based perception on a random demand, representing a surrogate measure of barrier integrity. 

Bayesian inference gives the formal procedure to update the priors into \emph{posterior} values of barrier integrity (and event probability), which  represent what the operational system \emph{currently} achieves. Thus, for our running example, the posterior integrity for $\barrierid{4}$ is given by (the proportional form of) Bayes' theorem as $\pi(p|y) \propto \mathcal{L}\left(p|n,y\right)\times \pi(p)$. Since the beta prior and the binomial likelihood are a conjugate pair, the posterior 
has a closed form solution, $\pi(p|y) \sim \mathtt{Beta}\left(s+y, (t-s)+x\right)$. The distribution mean, $\mu_{p|y} = \nicefrac{(s+y)}{(t+n)}$, is the updated point estimate of barrier integrity. To get a revised assessment of the operational safety risk level for the system, we propagate the posterior barrier integrity through the risk assessment model underlying the safety architecture.

\subsection{Characterizing the Change to Safety Risk}
\label{ss:rl-change-characterization}

We use \emph{risk ratio} (RR), a metric of relative risk, to quantify the change in operational safety risk. In operation, the RR for a consequence event is the ratio of its current estimated probability of occurrence and the approved baseline. More generally, we will (re)compute the RR for any event of interest in the safety architecture, typically after the operational risk assessment has been revised (as in Section~\ref{ss:update}) as the ratio of its updated (\ie prior or posterior, as appropriate) probability to its (scenario-specific) safety target. 
Denoting the RR for event $\eventid{i}$ by $\rr(\eventid{i})$, $\rr(\eventid{i}) >1$ indicates an increase in the safety risk of $\eventid{i}$. Similarly, $\rr(\eventid{i}) <1$ indicates a decrease, while $\rr(\eventid{i}) =1$ indicates no change. 
By itself, RR reflects how effective the safety architecture is in reducing the risk of the identified safety effects.\footnote{RR has also been used as a development safety metric, \eg in designing aircraft collision avoidance systems~\cite{daa-nmac-rr}.} By considering the trend of RR over time, we can construct a powerful SI of practical drift, \eg by fitting a linear trend line to a temporally ordered sequence of RR values computed over some pre-determined exposure, the sign and magnitude of the slope indicate, respectively, the direction and the rate of the change in safety risk.

\subsection{Numerical Examples}\label{ss:numerical-examples}

We now give some numerical examples to concretize the preceding discussion.

\begin{theorem}[Prior Barrier Integrity]\label{ex:prior}
During the development of our running example system and its pre-deployment safety case, assume we have a total of $\metric{devTxLowVisW}= 32$ flight tests in which there are $\metric{devPcpDisEngF} = 8$ failing tests for the \emph{Perception Failover} barrier. Thus a prior distribution for its integrity is $\pi(p) \sim \mathtt{Beta}(24, 8)$, whose mean is $\mu_{p} = 0.75$, and variance is $\sigma^{2}_{p} = 0.0057$. The mean gives a point prior value of barrier integrity which we show in the corresponding node in the BTD of Fig.~\ref{f:btd-example}.

\end{theorem}

\begin{theorem}[Scenario-specific Barrier Safety Indicator]\label{ex:barrier-spi}
Recall that a scenario-specific SI for the \emph{Perception Failover} barrier is $\spi{PFO}$: $\metric{opPcpDisEngF} \leq y$ in $n$ taxi operations (Section~\ref{ss:safety-indicators}). As before, $\metric{opPcpDisEngF}$ measures the number of failed disengagements of ML-based perception in operation. To determine a suitable exposure $n$ and threshold $y$, consider that a conservative range of values for $p$ that would provide the same, or better, risk reduction performance as its prior mean is the closed interval $\left[\mu_{p}+\sigma_{p}, 1\right] = [0.8254, 1]$. In other words, observing $8$ or more successful disengagements or, equivalently, $2$ or fewer failed disengagements on demand of ML-based perception over at least $10$ taxi operations conducted in the specified operating context would validate the safety performance of the barrier. Thus, here, $n = 10$ and $y = 2$.
\end{theorem}

\begin{theorem}[Likelihood of Data and Posterior Integrity]\label{ex:posterior}
After system deployment, suppose that to improve runway utilization, 
the control in $\barrierid{4}$ (see Fig.~\ref{f:btd-example}) is relaxed such that ML-based perception is disengaged after a larger distance (or duration) of position deviation than what was specified in the safety architecture. 
The metric that records the number of failed disengagements in operation, $\metric{opPcpDisEngF}$ (Section~\ref{ss:update}), includes violations of the barrier requirement as initially specified, which itself includes violations of the barrier requirement after operational modification. That is, the operational safety metric \emph{should not be modified} even though the barrier function has been operationally changed. Supposing $\metric{opPcpDisEngF} = 4$ violations have been observed over $\metric{opTxLowVisW} = 10$ taxi operations. 
Given this data, the likelihood function 
is $\mathcal{L}(p|6) = \binom{10}{4}p^{6}(1-p)^{4}$, and the posterior distribution is $\pi(p|6) \sim \mathtt{Beta}(30, 12)$, whose mean is $\mu_{p|6} = 0.7143$ and variance is $\sigma_{p|6}^{2}=0.0047$. 

\end{theorem}

\begin{theorem}[Operational Safety Risk Update]\label{ex:ora-update}
We assume prior data is available (from characterizing the ODD~\cite{kape-safecomp-2023} for the autonomous taxiing function) on how often runway markings are obscured during taxiing due to runway surface and weather conditions. Hence we can give a prior distribution, say $\pi(\eventid{2}) \sim \mathtt{Beta}(10, 190)$, whose mean is the prior point estimate $\mathtt{Pr}(\eventid{2}) = 0.05$. Similarly, let $\mathtt{Pr}(\eventid{1})=0.05$. Given these priors 
and the barrier integrity values as in Fig.~\ref{f:btd-example}, the prior probability of the consequence event is $\mathtt{Pr}(\eventid{4})= 1.5998\times 10^{-5}$ corresponding to an RRL of $\mathsf{4D (Low)}$.
We recall from Example~\ref{ex:posterior} that $4$ barrier violations were observed in $10$ taxi operations. Hence $\eventid{2}$ must have occurred on $z = 4$ occasions for $\barrierid{4}$ to have been invoked and have failed on demand. Thus, we may reasonably model this event as a Bernoulli process with a binomial PDF as the likelihood function for the observed data. Thus, the posterior distribution over $\eventid{2}$ is $\pi(\eventid{2}|z) \sim \mathtt{Beta}(14, 196)$ so that $\mu_{\eventid{2}|z} =  0.0667$ is the point posterior for $\mathtt{Pr}(\eventid{2})$. Propagating both the posteriors for $\eventid{2}$ and $\barrierid{4}$ through the risk assessment model of the safety architecture~\cite{dpw-ress2019}, we get the updated prior 
$\mathtt{Pr}(\eventid{4})= 2.386\times 10^{-5}$ for the consequence event. The corresponding RRL remains unchanged suggesting that the operational modification to $\barrierid{4}$ may be acceptably safe.

\end{theorem}

\begin{theorem}[Safety Risk Level Change and Practical Drift]\label{ex:rrl-change}
The risk ratio for the consequence event $\eventid{4}$ given the change to barrier $\barrierid{4}$ (as in Example~\ref{ex:posterior}) is $\rr(\eventid{4})=\nicefrac{2.386}{1.5998} \approx 1.49$. Thus, despite an unchanged risk level (see Example~\ref{ex:ora-update}), the RR indicates increasing safety risk. 
Now, further suppose that to improve runway utilization, a greater deviation in CTE from the stated bounds is operationally admitted (see the top event $\eventid{3}$ in Fig.~\ref{f:btd-example}). Consequently barrier $\barrierid{5}$ needs to be relaxed to be invoked after a longer duration than specified (see Fig.~\ref{f:btd-example}). 
Suppose that the posterior integrity computed from operational safety metrics (omitted here due to space constraints) is $\mathtt{Pr}(\barrierid{5}) \approx 0.96$. In this case, the revised prior for $\eventid{4}$ is $\mathtt{Pr}(\eventid{4}) \approx 2.4\times 10^{-4}$, the revised RRL is $\mathsf{4C (Medium)}$, and $\rr(\eventid{4})\approx 10$.
The updated RRL now violates the TLOS even if no safety effects may have been observed. Moreover, the modifications to the barriers $\barrierid{4}$ and $\barrierid{5}$ are at least \emph{an order of magnitude more likely} to result in a lateral runway overrun, indicating an appreciable increase in safety risk relative to the approved baseline, and suggests practical drift.

\end{theorem}

\section{Towards Formal Foundations}\label{s:formal-foundations}

As mentioned earlier (Section~\ref{ss:safety-measurement}), we want to formalize a notion of \emph{consistency} between the static portion of the safety case (\ie its assurance artifacts, see Fig.~\ref{f:dynamic-metamodel}) and the collection of indicators that constitute the \ac{smb}. The safety metrics and indicators represent the objectively quantifiable content referenced in the arguments and the safety architecture, which in turn provide the justification for how the metrics and indicators collectively provide safety substantiation. 
Although operational safety measurement entails updating and revising the risk assessment (Section~\ref{s:framework}), changing the SMB may not be necessary. However, in situations where replacing, modifying, or adding metrics and indicators is required---\eg to reflect new observable phenomena in the environment---the SMB will change and so would the associated assurance artifacts to retain consistency. Note that currently we are not considering changes that would entail modification of the safety architecture (\eg replacing a barrier). Hence we exclude that from our notion of consistency for now and focus on consistency with arguments. 

We can achieve this consistency if the argument structure reflects the risk reduction rationale implicit in the safety architecture. That is, the form of the argument structure proceeds from all terminating consequence events in the safety architecture, working recursively backwards (\ie leftwards) to all initiating (leftmost) threat events. Thus, each level of the argument has the following form: all consequence events are acceptably mitigated (\ie the residual risk level meets the allocated TLOS), \emph{which is supported by the argument that}:
all their identified precursor events (causes) are acceptably mitigated, \emph{which is supported by the argument that}:
\begin{inparaenum}[(a)]
	\item all applicable barriers are operational and effective, and
	\item all causes have the stated probability of occurrence.
\end{inparaenum}
In a GSN representation of this argument, the leaves are \emph{solution nodes}~\cite{gsn-std-v3} that have the following \emph{evidence assertion}: the initiating threat has the stated (assumed) probability. 

Thus, the overall argument states that if the barriers are effective and operational, and the events have the assumed probabilities, then the consequences have acceptable risk levels. 
Indicators map into the corresponding claims of barrier effectiveness and event probabilities, serving to monitor that those values are within the required limits.

Now we briefly outline how to place this consistency on a more rigorous basis. Let $\mathbf{Arg}$ and $\mathbf{SMB}$ represent the sets of well-formed arguments and \ac{smb}s, respectively, and define mappings $F : \mathbf{Arg} \to \mathbf{SMB}$ and $G : \mathbf{SMB} \to \mathbf{Arg}$, such that $F$ extracts the associated indicators from an argument, and $G$ embeds an \ac{smb} into a skeleton argument of the form outlined above. Then we require that $F ; G \leq I$ and $G ; F = I$ (where $I$ is the identity mapping), where arguments are ordered by \emph{refinement}. The first inequality ensures that the argument contains the necessary rationale for the \ac{smb}, with the refinement allowing that the argument can contain additional reasoning; the second ensures that all quantifiable components of the argument are represented in the \ac{smb}.

\section{Concluding Remarks}\label{s:conclusions}

\begin{figure}[t]
	\centering
	\includegraphics[width=\columnwidth]{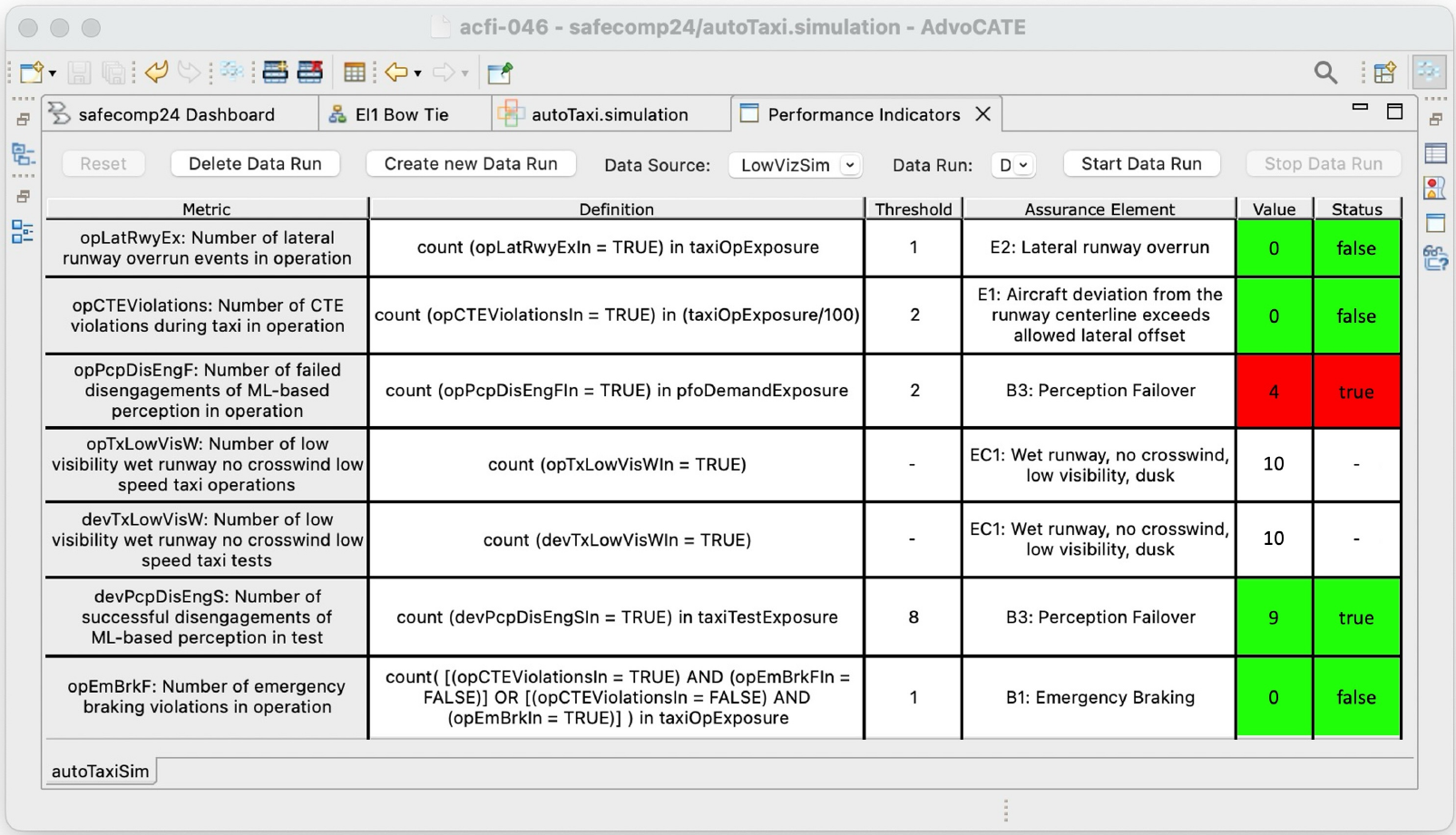}
	\caption{\acate screenshot: Table of safety indicators for the 
	example system in Section~\ref{s:example}.}
	\label{f:metrics-table}
\end{figure}

We have a preliminary implementation of the \ac{smb} in \acate that currently supports the following functionality: real-time import of data (\ie measures) from multiple data sources (simulations or feeds from external sensors); computation of derived metrics and indicators over multiple data runs; and  tracing to assurance artifacts (events and barriers in the safety architecture, and goals and assumptions in the safety arguments). 
We display indicators and the associated assurance artifacts in a dynamically updated table (Fig.~\ref{f:metrics-table} shows an example) that highlights when the conditions on the indicator thresholds have been met (in green) or have not been met (in red). 
A dashboard (not shown) allows selection between the various metrics of the \ac{smb} with charts displaying real-time updates of their values as well as other dynamically updated risk status, such as hazards ordered by risk level, and barriers ordered by integrity.

The goal of managing practical drift has mainly informed our choice of safety metrics and indicators. We plan to leverage the \emph{Goal Question Metrics} (GQM) approach~\cite{gqm} to define additional metrics suitable for other dynamic assurance goals, \eg improving functional performance whilst maintaining safety.

A binomial likelihood may be only initially appropriate for certain kinds of measurement data. Indeed, as more data is gathered, the preconditions for using a binomial PDF need to be reconfirmed. As such, it may be necessary to use other PDFs for the likelihood of the data, along with numerical methods for Bayesian inference. 
Our choice of beta priors is motivated, in part, by computational convenience, its flexibility to approximate a variety of distributions, and the domain-specific interpretation of the distribution parameters in different safety metrics. 
Although we represent the uncertainty in barrier integrity and event probability by specifying their distributions in the theoretical framework, our prototype implementation currently represents and propagates their point values (\ie the distribution means) for both the pre-deployment risk assessment, and the revisions of the operational risk assessment. We plan to refine this approach by also propagating the uncertainties through the risk assessment model so as to quantify the corresponding uncertainty in the residual risk of the safety effects of interest. By so doing, we aim to ground the quantification of assurance in safety measurement.

Since TLOS is typically assigned to rare events, legitimate concerns can arise about the credibility of using quantitative methods as in this paper. Though we have yet to explore how \emph{conservative Bayesian inference}~\cite{strigini-ivds21} could be used in our approach, relative risk metrics such as risk ratio (RR) are a step towards circumventing those concerns.

Practical drift is distinct from \emph{operating environment drift} in that the former results from changes within the system boundary, whereas the latter occurs outside that boundary. We reflect the assumptions about the operating environment in the pre-deployment safety case, for example, as the prior probabilities (conditional on the operating context) associated with the threat events. 
We can reflect environment drift via the posterior distributions of the corresponding event probabilities updated by operational safety metrics associated with the respective events (see Section~\ref{ss:update}, and Example~\ref{ex:ora-update}).
We additionally distinguish \emph{runtime risk assessment}~\cite{dac-computer}, from the update and revision of operational risk as described in this paper: the former occurs during the shorter time span of a mission (\eg during a taxi operation), whereas the latter occurs over longer time intervals, between missions, and through the lifecycle of the system (\eg over multiple taxiing operations, possibly involving an aircraft fleet).

The numerical examples (Section~\ref{ss:numerical-examples}) have described a scenario-specific application of our approach, where the event probabilities and barrier integrities are \emph{conditional} on the operating context. For a system-level characterization of how operational risk changes, we must consider the \emph{marginal} probabilities and integrities in the overall safety architecture that composes different risk scenarios. However, we have not considered it in this paper, and it is one avenue for future work. 

To further develop our proposed dynamic assurance framework we aim to explore how by thresholding, ranking, and comparing RR under changes made to individual mitigations or their combinations, we may infer: 
\begin{inparaenum}[(i)]
	\item which mitigations may be optimized for system performance whilst maintaining safety (possibly necessitating a change to the safety architecture itself); and, in turn, 
	\item which system and safety case changes may be necessary. Some changes may be automated while will induce \emph{tasks} requiring manual attention \cite{dhp-icse-nier-2015}.
\end{inparaenum}
Additionally, we aim to define a tool-supported methodology on top of the main components of the framework. This will involve defining and formalizing the methods and procedures to decompose and allocate safety targets, derive safety indicators, and close the safety assurance loop, \ie maintain consistency of the arguments with the SMB) through targeted changes to the system and its safety case.

Observations of system operations constitute one specific form of evidence that we can use to reason about system safety. We seek to systematize this through a notion of \emph{evidence requirement} that will also cover \emph{static} data. We are also extending the metrics expression language to express trends, although work remains to integrate it into our methodology and to relate it to the concept of \emph{safety objective}. A need to update the \ac{smb}, \eg modify indicators and possibly their thresholds, accompanies operational safety measurement. We aim to better understand the principles that underlie those modifications and, subsequently, implement the corresponding tool features.  
However, practically deploying this framework will necessitate harmonizing with existing safety management system (SMS)~\cite{faa-ato-sms} infrastructure, whilst carefully considering the roles of different stakeholders in safety performance monitoring, measurement, and assurance.

\bibliographystyle{splncs04}

\end{document}

\typeout{get arXiv to do 4 passes: Label(s) may have changed. Rerun}